\documentclass[cits]{PoS}
\usepackage{amsmath}
\usepackage{wrapfig}

\def\be{\begin{equation}}
\def\ee{\end{equation}}
\def\bea{\begin{eqnarray}}
\def\eea{\end{eqnarray}}
\def\({\left (}
\def\){\right )}
\def\[{\left [}
\def\[{\right ]}
\newcommand{\Tr}{{\rm Tr}}

\title{Holographic thermalization of mutual and tripartite information in 2d CFTs}

\ShortTitle{Holographic thermalization}

\author{Alice Bernamonti\\ %
 Instituut voor Theoretische Fysica, KU Leuven,\\Celestijnenlaan 200D, B-3001 Leuven, Belgium  \\
        E-mail: \email{alice@fys.kuleuven.be}}
\author{Neil Copland\\
Institute for Theoretical Physics, Masaryk University,\\ 611 37 Brno, Czech Republic\\
        E-mail: \email{ncopland@mail.muni.cz}}
\author{Ben Craps\\
Theoretische Natuurkunde, Vrije Universiteit Brussel, and International Solvay Institutes,\\Pleinlaan 2, B-1050 Brussels, Belgium \\
        E-mail: \email{ben.craps@vub.ac.be}}
\author{ \speaker{Federico Galli}\\
Theoretische Natuurkunde, Vrije Universiteit Brussel,  and International Solvay Institutes,\\Pleinlaan 2, B-1050 Brussels, Belgium \\
        E-mail: \email{federico.galli@vub.ac.be}}

\abstract{We discuss the concepts of mutual and tripartite information by referring to simple intuitive examples, and describe their use as probes of thermalization in holographic models. We then review our computation of these quantities in a simple time dependent model, in which energy injection in a strongly coupled field theory is modeled by a shell of null dust falling into 3d anti-de Sitter space. We complete those results with a discussion of the possible equilibrium phases of the tripartite information.}

\FullConference{ Proceedings of the Corfu Summer Institute 2012 \\

                 September 8-27, 2012\\

                 Corfu, Greece}

\begin{document}

\section{Introduction}

Applications of the AdS/CFT correspondence to particle and condensed matter physics have been a subject of intense research for a number of years now. A recent development in this area is the study of non-equilibrium processes in strongly coupled field theories, in particular the approach of thermal equilibrium starting from far-from-equilibrium initial conditions. 

One motivation is the creation of a quark gluon plasma in heavy ion collisions. The observed anisotropy in the distribution of particles in the final state points to a collective flow, which can be described by hydrodynamical simulations. To match the data, one needs the viscosity to be small, which implies that the plasma is strongly coupled. Moreover, the hydrodynamic regime has to start early enough and last long enough. The early onset of a hydrodynamic regime, earlier than would be expected based on weak coupling estimates, is often referred to as fast thermalization. In the absence of other techniques to study strongly coupled systems out of equilibrium, it is natural to turn to holography to try and develop an understanding of this fast thermalization process, i.e., of the way the system reaches local thermal equilibrium. 

Other systems of interest are quantum quenches, the response of a system to a time-dependent coupling. Many such simple models can be realized experimentally in cold atom systems, and powerful theoretical techniques have become available to analyze them. In particular, Calabrese and Cardy have obtained many results on quantum quenches from gapped systems to 2d conformal field theories, including  the behavior of entanglement entropy in these models \cite{Calabrese:2005in}. We will find it useful to compare our results on holographic thermalization models to theirs on quantum quenches.

According to the AdS/CFT correspondence, a thermal state in the field theory is dual to a black brane in anti-de Sitter (AdS) space. A thermalization process in field theory therefore corresponds to the formation of a black brane in AdS. Hence, to quantify how a field theory thermalizes as a function of time we will study the time evolution of a black brane formation process. 

This clearly requires probes that are localized in time. The simplest probes one can think of are expectation values of gauge invariant operators, which we refer to as one-point functions. However, in the model we will study, these assume thermal values immediately after a sudden homogeneous injection of energy, as if the system had thermalized instantaneously. To show that the thermalization process is not instantaneous, one needs non-local probes. One class of such probes are two-point functions,  as discussed in Joris Vanhoof's talk at this workshop. Here we will be interested in probes related to entanglement entropy, namely the mutual and tripartite information. 

Before giving technical definitions, let us briefly characterize the mutual information $I(A,B)$ of two systems $A$ and $B$. It measures how much information the systems share, answering the question: What do we learn about $A$ by looking at $B$? It is non-negative, and is zero only if the two systems are uncorrelated.

At the intuitive level we can think of  a simple example where $A$  and $B$ are two books. If $A$ and $B$ are two different editions of the same book, reading $A$ teaches us basically everything about the content of $B$. On the other hand if  $A$ and $B$ are two books that have no relation between each other, for instance a novel and a handbook of integrals, there is no information that can be gained about one book by reading the other. This situation would be basically characterized by a vanishing mutual information, while the former  would correspond to a large value of $I(A,B)$. 

The tripartite information can be defined in terms of the mutual information as
\be
I_3(A,B,C)=I(A,B)+I(A,C)-I(A, B\cup C), \label{TI}\,
\ee
which turns out to be symmetric under the interchange of $A$, $B$ and $C$, and it quantifies how much information $A$ shares with $B$ and $C$ separately, relative to what it shares with $B$ and $C$ together. The tripartite information can take any sign. If $I_3\leq 0$, $I(A, B\cup C)$ constitutes an upper bound for the information that $A$ can share with $B$ and $C$ separately -- in this case, the mutual information is said to be ``monogamous''. 

\begin{figure}[h]
\centering

\includegraphics[width=0.7\textwidth]{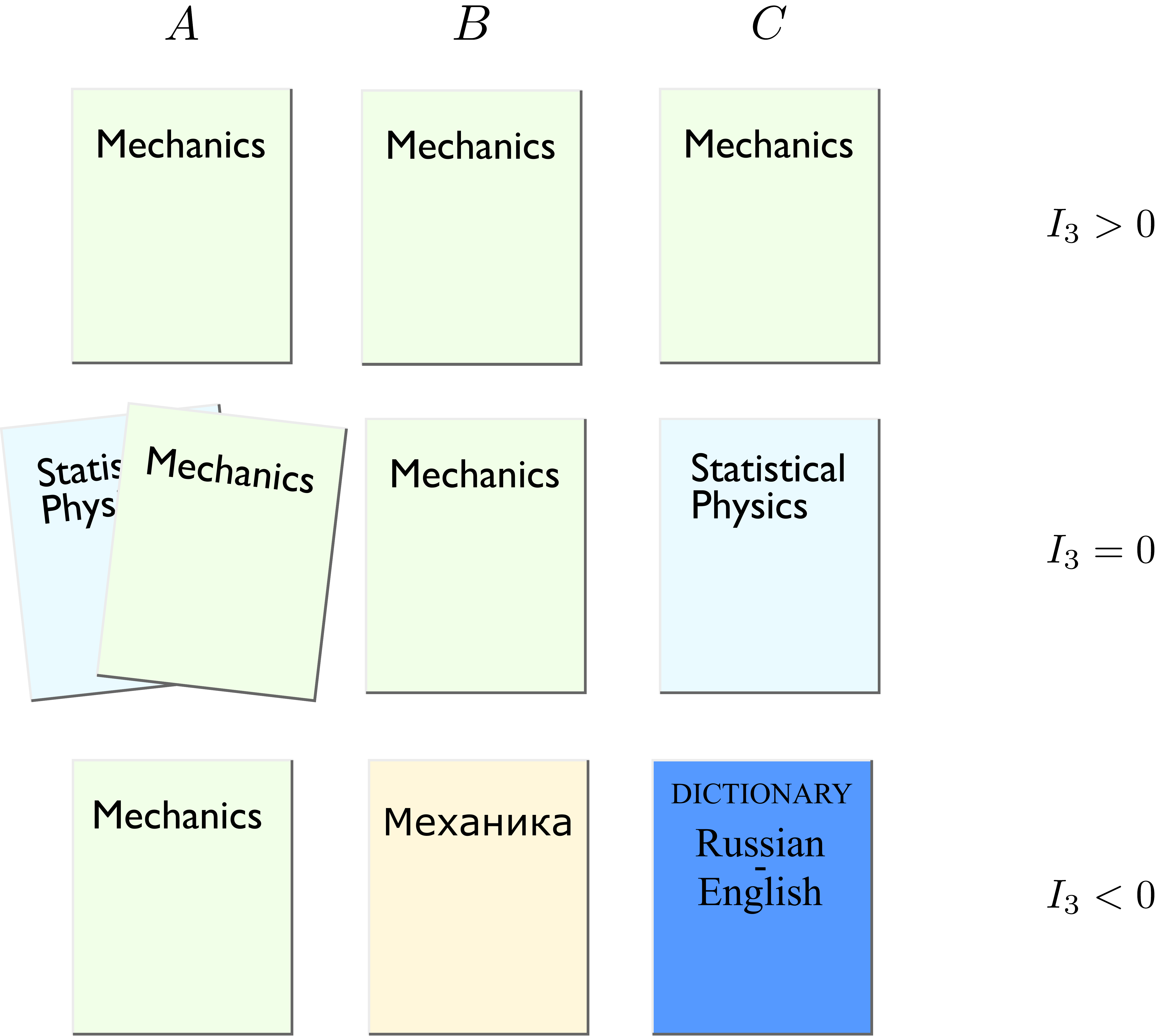}
\caption{A simple example that gives some intuition about the tripartite information. As illustrated in the three  different cases, the tripartite information can take any sign.}
\label{fig:TIex}
\end{figure}

One can get some intuition on this quantity using again the example where $A$, $B$ and $C$ are three books.  For three copies of the same book, there is a redundancy in the information that $A$ shares with $B$ and $C$. This corresponds to a positive tripartite information, as illustrated in the first line of Fig.~\ref{fig:TIex}. In the second line of the same figure, the situation where  $B$ and $C$ are two different books is depicted.  Reading $B$ gives  partial information about $A$, while reading $C$ would give the complementary piece of information.  Such a situation thus corresponds to a vanishing tripartite information. Finally, in the third line of Fig.~\ref{fig:TIex}, $B$ is the Russian version of the English book $A$, while $C$ is a Russian-English dictionary. Having access exclusively to $B$ or to $C$, would not teach us anything about the content of $A$ (unless we were able to read Russian). However the system $B \cup C$ allows to know the entire content of $A$ thus resulting in a negative tripartite information.
%

\section{Setup}

\begin{figure}[h]
\centering
\includegraphics[width=0.4\textwidth]{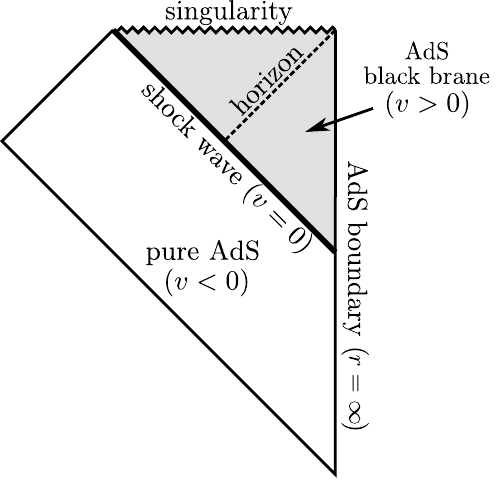}
\caption{The causal structure of the thin-shell AdS$_3$-Vaidya spacetime in the Poincar\'e patch.  The planar AdS boundary corresponds to the vertical line on the right, while the Poincar\'e horizon is given by the null lines on the left. }
\label{fig:vaidyapenrose}
\end{figure}

We focus on a very simple holographic model, in which a sudden but homogeneous injection of energy in a 2d conformal field theory is modeled by a shell of null dust falling into 3d AdS. The resulting spacetime is described by  the thin-shell AdS$_3$-Vaidya metric
\be\label{eq:Vaidya}
ds^2= -[r^2-r_H^2\,\theta(v)]\, dv^2+2drdv+r^2 dx^2,
\ee
where $\theta$ is the  step function, the AdS boundary is located at $r \to \infty$ and the shell of null dust at $v=0$ (see Fig.~\ref{fig:vaidyapenrose}). Inside the shell, $v<0$, the spacetime reduces to pure AdS in ingoing Eddington-Finkelstein coordinates. Outside the shell, $v>0$, the spacetime is that of an AdS black brane with temperature $T_H=r_H/(2\pi)$.

The probes we will be discussing are closely related to entanglement entropy. Consider a quantum system described by a density matrix $\rho$, and divide it in two complementary subsystems $A$ and $B$. (These could correspond to complementary spatial regions in a quantum field theory, for instance.) When considering observables in $A$ only, we can trace over the degrees of freedom in $B$ to obtain the reduced density matrix $\rho_A=\Tr_B(\rho)$. The entanglement entropy of $A$ is then defined as the von Neumann entropy of this reduced density matrix,
\be
S(A)=-\Tr_A(\rho_A\log\rho_A).
\ee
If the total density matrix $\rho$ is that of a pure state, then $S(A)=S(B)$, and the entanglement entropy measures to what extent the degrees of freedom in $A$ are entangled with those in $B$.

A benefit of working with the entanglement entropy is that a simple proposal exists for computing it in holographic models. Let us focus first on a spatial interval $A$ (at an arbitrary moment of time $t_0$) in a 2d CFT in an equilibrium state. According to Ryu and Takayanagi \cite{Ryu:2006bv}, the relevant quantity to compute is the length of the minimal bulk geodesic $\gamma_A$ in the $t=t_0$ plane that connects the endpoints of the boundary interval $A$. The entanglement entropy of $A$ is then proposed to be
\be \label{eq:S}
 S(A)=\frac{{\rm length}(\gamma_A)}{4G_N}.
\ee
An extension to non-equilibrium states has been proposed in \cite{Hubeny:2007xt}. In that case, \eqref{eq:S} generalizes to extremal geodesics that will generically not lie in an equal-time surface. 

\begin{figure}[t]
\centering
\includegraphics[width=\textwidth]{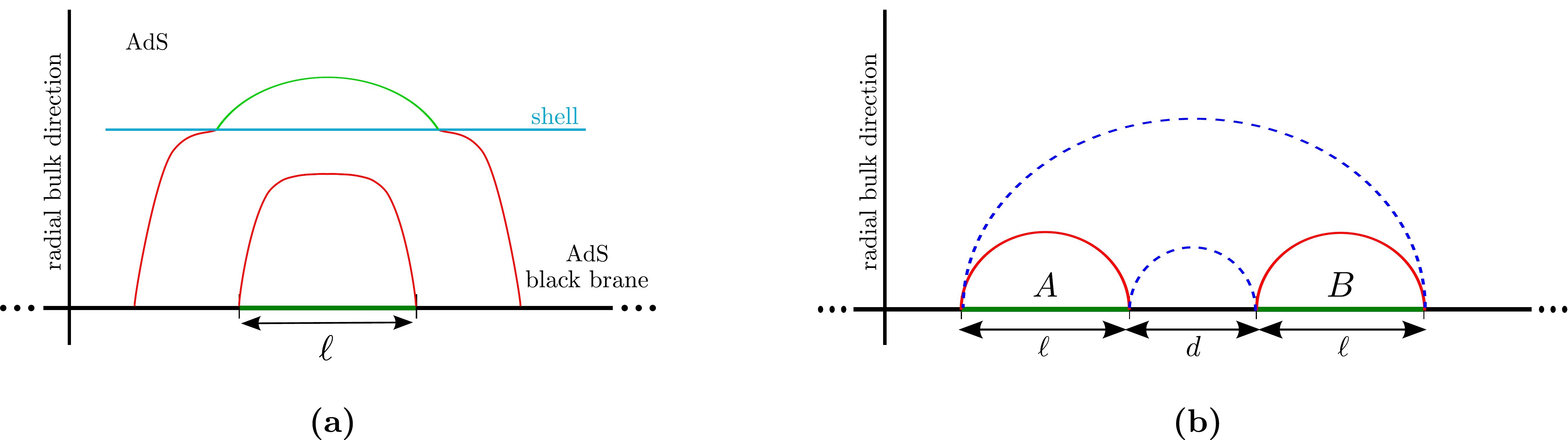}
\caption{({\bf a}) A schematic representation of the possible configurations for fixed boundary time geodesics in the infalling shell background. We remark that this is just a cartoon and that configurations that cross the shell will not lie on equal time surfaces in the bulk. ({\bf b}) Competing geodesics configurations for the entanglement entropy $S(A \cup B)$ of two disjoint intervals of length $\ell$ on a fixed time surface in AdS$_{3}$. Depending on the separation  $d$  the connected configuration in blue dashed lines or the disconnected configuration  in red continuous lines will be the minimal one.} 
\label{fig:EEMIholo}
\end{figure}

The specific case of AdS$_3$-Vaidya spacetime was studied in \cite{AbajoArrastia:2010yt,Balasubramanian:2010ce}. For negative values of the fixed boundary time $t_0$ ($v_0 \equiv t_0$ on the AdS boundary) the full extent of these geodesics is within the pure AdS geometry. The entanglement entropy then takes its vacuum equilibrium value\footnote{The geodesics length ${\mathcal L}$ is divergent in asymptotically AdS spacetime, due to boundary contributions. It is more convenient to work with a renormalized geodesics length  ${\delta \mathcal L}$ obtained introducing a large radius cutoff $r_{0}$  and subtracting the divergent part: $\delta \mathcal L \equiv \mathcal L - 2 \ln 2 r_{0} $. }%
\be \label{vacuum}
\delta S_{\text{vacuum}} = \frac{c}{3} \ln \frac{\ell}{2}\,,
\ee
where $c=2/(3G_{N})$. 
When $t_0\geq0$ different situations are possible depending on the separation $\ell$ between the endpoints of $A$;  a schematic representation is given in Fig.~\ref{fig:EEMIholo}a. For small enough separation, that is $\ell \leq 2 t_0$, the geodesics lies entirely in the AdS black brane geometry. This corresponds to situations where the entanglement entropy assumes its thermal value
\be \label{thermal}
\delta S_{\text{thermal}} = \frac{c}{3} \ln \frac{\sinh \frac{r_{H}\ell}{2}}{r_{H}}.
\ee 
Increasing the separation $\ell$, these geodesics will begin at the boundary, extend into the AdS black brane geometry, cross the shell with a suitable refraction condition and continue into the pure AdS geometry from where they return to the boundary crossing again the shell. An analytic expression for the length of this kind of geodesics has been obtained in \cite{Balasubramanian:2010ce}.\\

The mutual information between  two disjoint subsystems $A$ and $B$ is defined in terms of the entanglement entropy as  
\be \label{mutual}
I(A,B) = S(A) +S(B) - S(A \cup B)\, . 
\ee
It is a non-negative quantity and it vanishes if and only if there is no correlation between the two subsystems, i.e. $\rho_{AB} = \rho_A \otimes \rho_{B}$.
Compared to the entanglement entropy, it has the advantage of being finite; the divergences that are present in the entanglement entropy cancel in the mutual information of two disjoint intervals. 

In the case of a 2d CFT in an equilibrium state, it is straightforward to compute the mutual information holographically.  In light of the above discussion, the only missing ingredient is the entanglement entropy of the union of disjoint intervals. One can extend the equilibrium prescription to this case by selecting the minimal length configuration amongst all possible collections of geodesics connecting the endpoints of the disjoint intervals \cite{Headrick:2010zt}.  We assume this holds also in the non-equilibrium setup. 

Two disjoint intervals of length $\ell$ on the boundary of pure AdS spacetime are depicted in Fig.~\ref{fig:EEMIholo}b. Depending on the separation $d$ between the intervals and for a fixed value of $\ell$, the connected configuration in blue dashed lines or the disconnected configuration in red continuous lines will be the minimal one. In particular, the disconnected configuration dominates for large enough $d/\ell$. The configuration where geodesics cross is instead always suboptimal. The analog is true also in  AdS$_3$-Vaidya spacetime, with the important difference that geodesics can generically cross the shell and will thus not lie on equal-time surfaces \cite{Balasubramanian:2011at}.

Finally we recall the definition \eqref{TI} of the tripartite information between three subsystems $A$, $B$ and $C$
\be \label{eq:I3}
I_3(A,B,C)=I(A,B)+I(A,C)-I(A, B\cup C)\, ,
\ee
and that generically it can take any sign. Notice that the tripartite information computed in perturbative quantum field theory is always non-negative \cite{Balasubramanian:2011wt}, while general geometric considerations allow to prove that the mutual information is monogamous ($I_3 \leq 0$) in static holographic setups \cite{Hayden:2011ag}.

%
\section{Holographic probes}

Before considering the evolution of the mutual and tripartite information after a sudden injection of energy, let us review the results for the entanglement entropy  obtained in \cite{AbajoArrastia:2010yt,Balasubramanian:2010ce}.  

\begin{figure}[h]
\centering
\includegraphics[width=0.55\textwidth]{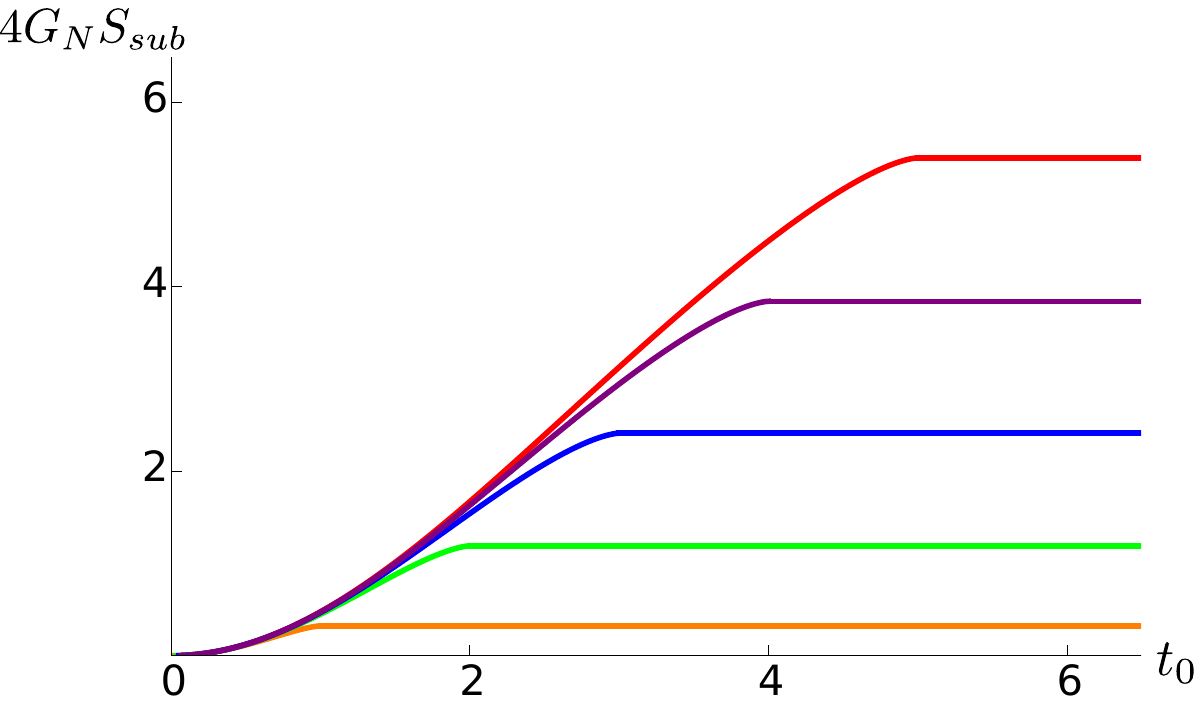}
\caption{The evolution of the entanglement entropy for an interval of length $\ell$ as a function of the boundary time. The vacuum  AdS value of the entanglement entropy has been subtracted and $r_{H}=1$.  The different curves corresponds to different values of $\ell =2, 4, \dots 10$ increasing from the bottom up.  }
\label{fig:EEAdSrescaled}
\end{figure}

The entanglement entropy of a single interval computed in the holographic setup \eqref{eq:Vaidya} starts from its vacuum value and increases almost linearly to reach its thermal value. This is shown in Fig.~\ref{fig:EEAdSrescaled}, where the evolution of the entanglement entropy for different values of $\ell$ is plotted. One sees that the thermalization time for an interval of length $\ell$ is $t_{th} = \ell/2$. The resulting thermalization dynamics is therefore top-down, meaning that  it proceeds from UV to IR modes. 

The main features of the evolution -- the intermediate linear regime and the thermalization time -- closely resemble the results obtained by Calabrese and Cardy for a system following a global quench to a 2d CFT \cite{Calabrese:2005in}.  
They considered a 2d quantum field theory with a mass gap in its ground state.  At some time, $t=0$, the mass gap is removed (for instance by suddenly changing  some tunable parameter in the Hamiltonian) leaving a 2d CFT in an excited state. Calabrese and Cardy  computed the resulting evolution of the entanglement entropy  in the limit where all lengths and time scales involved are large compared to the initial inverse mass gap \cite{Calabrese:2005in}.

\begin{figure}[h]
\centering
\includegraphics[width=\textwidth]{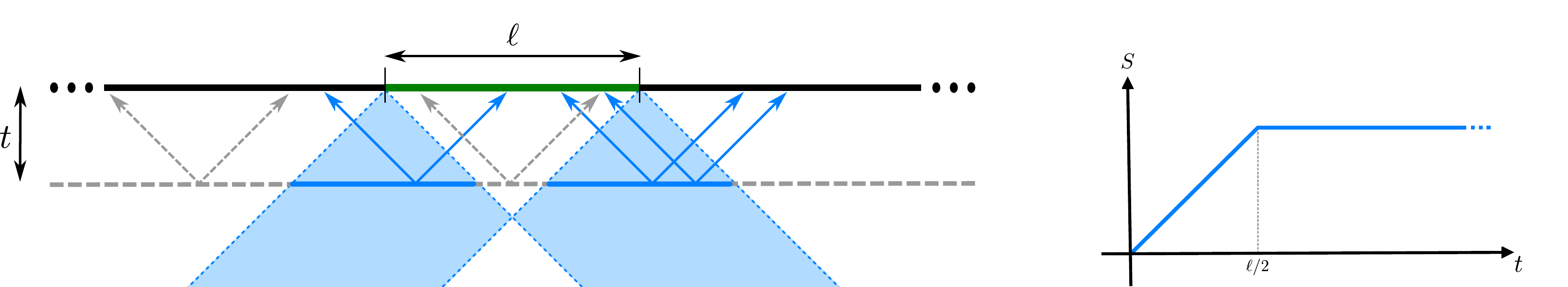}\\
\caption{The causal picture for the entanglement entropy. At some time $t$ after the quench, excitations originating from the two intervals in  blue continuous line will contribute to the entanglement entropy. For different time separation only excitations coming from the blue shaded region will contribute producing the time profile plotted on the right.}
\label{fig:causalEE}
\end{figure}

A simple toy model captures the physics after the quench \cite{Calabrese:2005in}. The idea is that after the quench there is a sea of quasi-particle excitations which are free to travel to the right and to the left at the speed of light. Because of the initial mass gap, two excitations are entangled only if they originate from the same point at the time of the quench. They contribute to the entanglement entropy of one interval with the complementary region only when one excitation is inside the interval and the other outside. At some instant of time $t$ after the quench, one can therefore quantify the entanglement entropy by counting these correlated pairs. Fig.~\ref{fig:causalEE} sketches this argument;  the blue continuous arrows correspond to a sample of the excitations that contribute to the entanglement entropy. This is measured as the length of the two intervals from which these quasi-particles originate.
Following its time evolution, we see that the entanglement entropy increases linearly and reaches its final equilibrium value in a time $t= \ell / 2 $.

We now discuss our probes: the mutual and tripartite information. The mutual information generically starts from the vacuum  AdS$_{3}$ value and interpolates to the thermal one, computed in the AdS$_{3}$ black brane background. These equilibrium values in general depend on the length of the two intervals considered and on their separation. For two intervals of equal length $\ell$ separated by a distance $d$ in the thermal black brane background  \cite{Tonni:2010pv}
\be\label{MIBTZ}
I(A,B)_{\text{thermal}} = \left\{\begin{array}{cc}0 \,,&\quad d \ge d_{\text{1th}}  \\
     					  \frac c 3 \ln \left[\frac{\sinh^2 (r_H \ell /2) }{\sinh(r_H (2 \ell+d)/2) \sinh(r_H d/2)}\right] \,,& \quad d \le d_{\text{1th}},
					 \end{array}\right.
\ee
 where 
\be \label{dthermal}
d_{\text{1th}}(\ell) = \frac{1}{r_H} \ln \left[ 1-e^{r_H \ell} + e^{2 r_H \ell} + \sqrt{(1-e^{r_H \ell} + e^{2 r_H \ell})^2 - e^{2 r_H \ell}} \right] - 2\ell \, .
\ee
In the $r_H \to 0$ limit, one recovers the vacuum AdS result of \cite{Headrick:2010zt}, 
\be
I(A,B)_{\text{vacuum}} = 
\left\{\begin{array}{cc}0 \,,&\quad  d \ge d_{\text{1v}}  \\
     					  \frac c 3 \ln \left[\frac{\ell^2}{d(2\ell+d)}\right] \,,& \quad d \le d_{\text{1v}},
					 \end{array}\right.
\label{MIAdS}
\ee	
with $d_{\text{1v}}(\ell) = (\sqrt 2 -1)\ell$.

\begin{figure}[]
\includegraphics[width=0.45\textwidth]{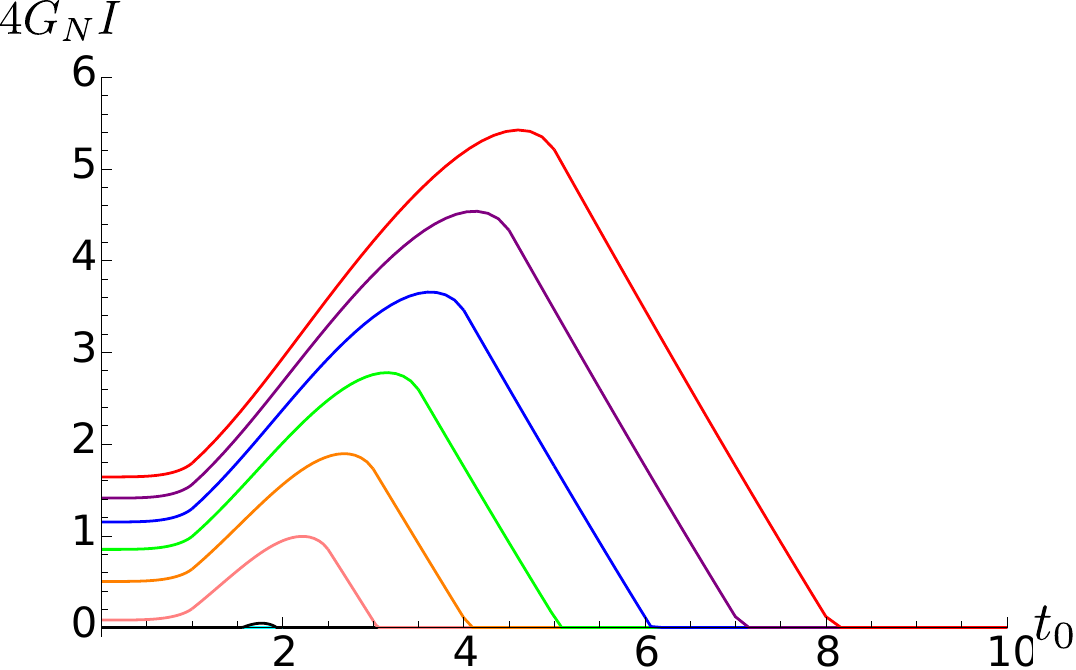} \hfill
 \includegraphics[width=0.45\textwidth]{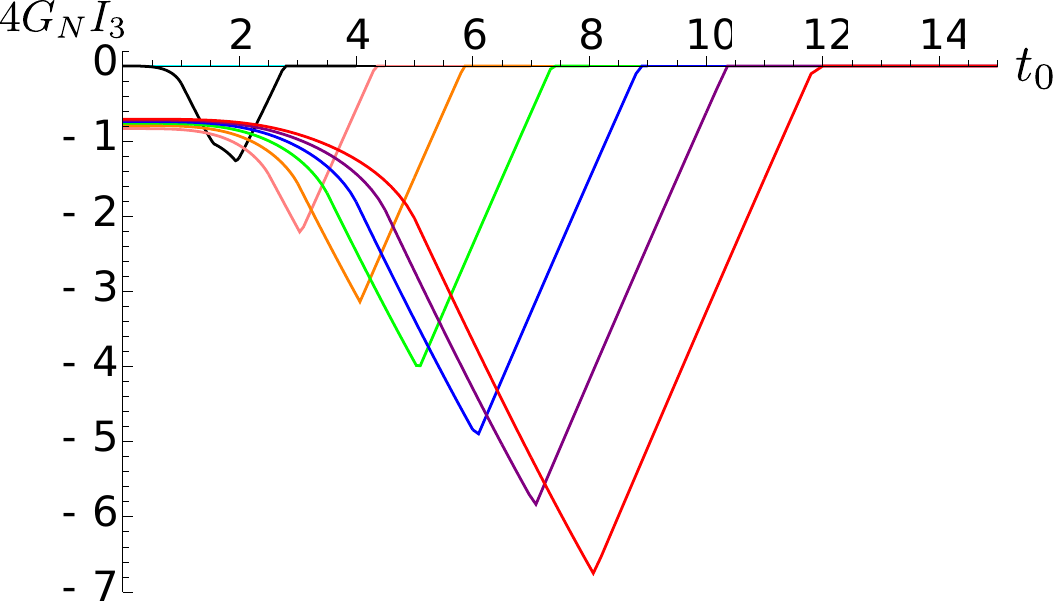}
\caption{ On the left panel, the mutual information for two intervals of equal length $\ell$ separated by a distance  $d =2$. The various curves correspond to $\ell = 1, 2,\dots 10$ increasing from the bottom up.  Some of the curves are not visible because they are everywhere vanishing. The right panel shows the tripartite information for three intervals of equal length $\ell$  separated by a distance $d =2$. The various curves correspond to $\ell = 1, 2,\dots 10$ decreasing from right to left, however, only the curves  corresponding to $\ell = 3,\dots 10$ are visible because the others are everywhere vanishing. }
\label{fig:MITI}
\end{figure}

However, the main feature in the evolution of the mutual information is independent of these details. For two disjoint intervals it has a distinct peak: rather than interpolating monotonically from the starting equilibrium value to the final one, it increases almost linearly, reaches a maximum and then decreases linearly, as shown in the left panel of Fig.~\ref{fig:MITI}.

\begin{figure}[ht]
\centering
\includegraphics[width= 0.9 \textwidth]{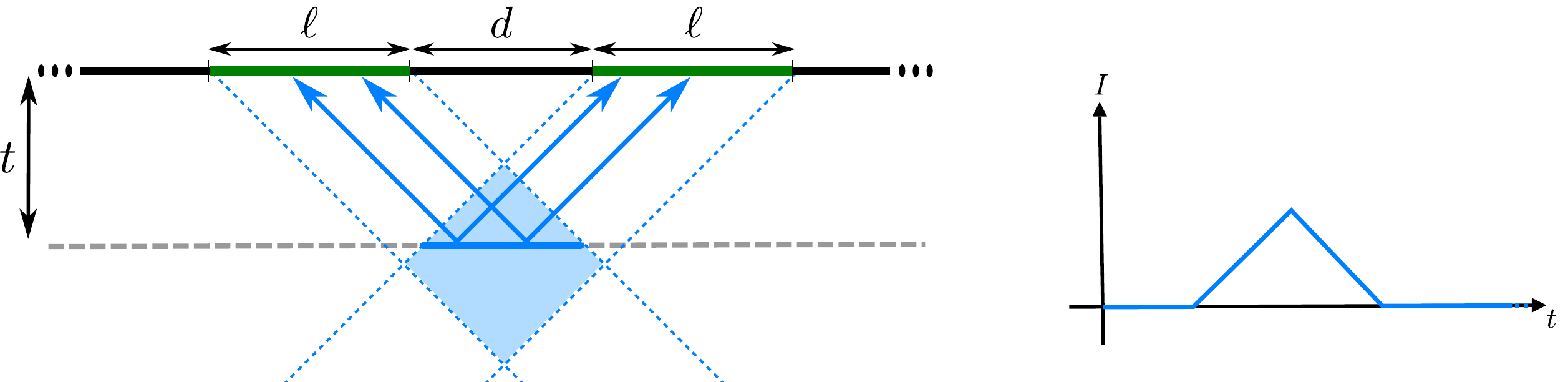}
\caption{The causal picture for the mutual information between two disjoint intervals of equal length. At some time $t$ after the quench, excitations originating from the blue continuous  interval will contribute to the mutual information that can be measured as the length of this interval. For different time separation, only excitations coming from the blue shaded diamond region contribute. The corresponding evolution of the mutual information is sketched on the right.}
\label{fig:causalMI}
\end{figure}

This behaviour turns out to be unexpectedly well captured by the simple causal argument reviewed above. At some time after the quench, correlated excitations originating from the same point at the moment of the quench will contribute to the mutual information between two intervals only if  they are each in a different interval. Therefore only pairs of quasi-particles originating from the blue shaded region in Fig.~\ref{fig:causalMI}  will contribute to the mutual information.
The corresponding evolution thus nicely reproduces the intermediate peak that we find in the holographic computation.

It seems therefore natural to try investigate how far this matching can be pushed by examining the tripartite information. The right panel of Fig.~\ref{fig:MITI} gives a prototypical example of the evolution of the tripartite information, for three disjoint intervals of equal length $\ell$, equally spaced from each other by a distance $d$.

\begin{figure}[ht]
\begin{center}
\includegraphics[width=0.5\textwidth]{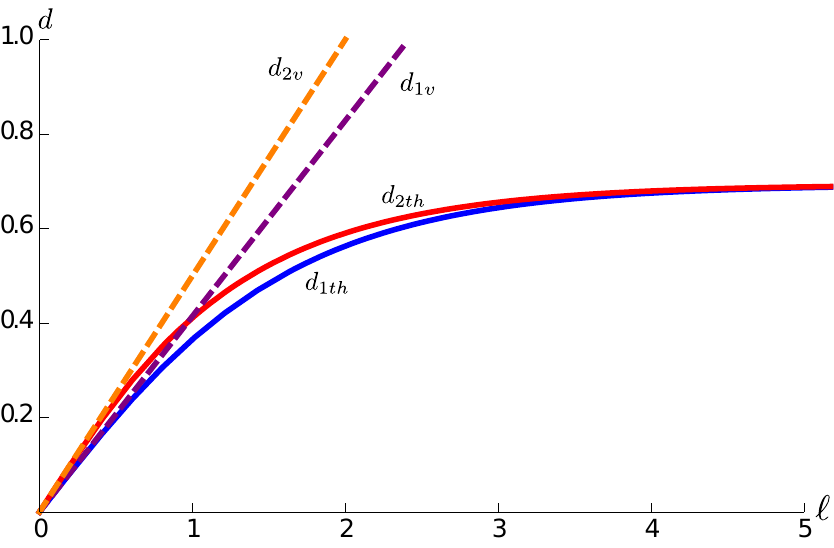}
\end{center}
\caption{Different phases of the equilibrium tripartite information  in the ($\ell$,$d$) plane for $r_H=1$. $d_{\text{1th} }$ and $d_{\text{2th}}$ are depicted in blue and red respectively, $d_{\text{1v}}$ and $d_{\text{2v}}$ in dashed purple and orange. Observe that  the bounds $d_{\text{1th} }$ and $d_{ \text{1v} }$ appear also in the mutual information, but that the tripartite information exhibits a third new phase delimited by $d_{ \text{2th} }$ and $d_{\text{2v}}$ in the thermal and vacuum case, respectively.}
\label{fig:tripartitethermal}
\end{figure}

Before discussing the non-equilibrium results, we include here for completeness the equilibrium result of the tripartite information \cite{alice}, which was not present in \cite{Balasubramanian:2011at}.
The tripartite information \eqref{eq:I3} in the thermal background is computed similarly to the mutual information, using \eqref{thermal} and \eqref{mutual}. 
The expectation value of this probe shows three different phases in the ($\ell$,$d$) plane (see Fig.~\ref{fig:tripartitethermal}):
\be \label{TIBTZ}
I_3(A,B,C)_{\text{thermal}} =  \left\{\begin{array}{cc} 0  \,, & d \ge d_{\text{2th}}  \\
     					  \frac c 3 \ln \left[\frac{\sinh (r_H (3 \ell +2d) /2) \sinh^2(r_H d/2) }{\sinh^3(r_H  \ell /2)}\right] \,,&  d_{\text{1th}} \le d \le d_{\text{2th}} \\
					  \frac c 3 \ln \left[\frac{\sinh (r_H(3 \ell +2d)/2) \sinh(r_H \ell /2))}{\sinh^2(r_H (2 \ell+d)/2)}\right] \,,&  d \le d_{\text{1th}},
					 \end{array}\right.
\ee
where $d_{\text{1th}}$ is defined in \eqref{dthermal} and 
\be
d_{\text{2th}} (\ell) = \frac{1}{r_H} \ln \left[ 1-e^{-r_H \ell} + e^{- r_H \ell}  \sqrt{1-e^{r_H \ell} + e^{2 r_H \ell} } \right] \,.
\ee
The vacuum result obtained from \eqref{TIBTZ} in the $r_H \to 0$ limit reads 
\be \label{TIAdS}
I_3(A,B,C)_{\text{vacuum}} =  \left\{\begin{array}{cc} 0  \,, & d \ge d_{\text{2v}}  \\
     					 \frac c 3 \ln \left[\frac{(3 \ell +2d) d^2}{\ell^3} \right] \,,&  d_{\text{1v}} \le d \le d_{\text{2v}} \\
					  \frac c 3 \ln \left[\frac{(3 \ell +2d)\ell }{(2 \ell +d)^2}\right] \,,&  d \le d_{\text{1v}},
					 \end{array}\right.
\ee
where $d_{\text{1v}} = (\sqrt 2 -1)\ell $ and $d_{\text{2v}} = \frac \ell 2$.{} \\

Notice that in the non-equilibrium situation, other profiles of $I_3 $ than those presented in Fig.~\ref{fig:MITI} are possible, depending on the separation and the length of the intervals. For instance, in the limiting case of adjacent intervals, we observe a plateau at an intermediate stage of the evolution \cite{Balasubramanian:2011at}. 
However, a common lesson can be drawn from these different cases.
The tripartite information is generically time dependent (it is constant only when it is everywhere vanishing). This contrasts with the results obtained by Calabrese and Cardy for the global quench. A simple calculation in fact shows that in that case the tripartite information is time independent. This can be intuitively understood in terms of the causal argument: it is not possible to populate three intervals with pairs of excitations. The result for the tripartite information thus clearly  shows that a free quasi-particle picture is not sufficient to fully take into account the strongly coupled dynamics and that interactions should be included. Notice that there is however no contradiction between the quench result and the holographic one. The two settings are in fact quite different from the outset. In the first case the system has short range correlations, while in the second one starts with long (actually infinite) range correlations from the beginning.

Moreover, as in the cases depicted in Fig.~\ref{fig:MITI},  we always found that during its evolution the tripartite information is non-positive. Therefore, the mutual information in our strongly coupled 2d thermalization scenario is monogamous, matching the result \cite{Hayden:2011ag} for strongly coupled theories at equilibrium that admit a gravity dual.

After \cite{Balasubramanian:2011at} came out, a number of interesting closely related results appeared in \cite{Allais:2011ys,Callan:2012ip,Wall:2012uf}. In \cite{Allais:2011ys,Callan:2012ip}  for AdS-Vaidya spacetime, and in \cite{Wall:2012uf} for other more general non-equilibrium configurations, it was shown that the holographic prescription of \cite{Hubeny:2007xt} used here satisfies the strong subadditivity condition. This is an important result in support of the validity of \cite{Hubeny:2007xt}. It is interesting to note that the key assumption in these works is that the null energy condition holds, which in \cite{Allais:2011ys} and  \cite{Wall:2012uf} was demonstrated to be a necessary condition for the monogamy of the mutual information.
\acknowledgments
FG would like to thank the organizers of the Corfu Summer Institute for the stimulating environment and nice atmosphere, and the organizers of the XVIIIth European Workshop on String Theory  for the opportunity to present  this work.
We would also like to thank Vijay Balasubramanian for collaboration on most of  the results on which these proceedings are based.
AB, BC and FG are supported in part by the Belgian Federal Science Policy Office through the Interuniversity Attraction Pole P7/37.
AB is supported in part by the FWO-Vlaanderen, Project No. G.0651.11 and in part by the European Science Foundation Holograv Network. AB is a Postdoctoral Researcher FWO-Vlaanderen.
NC is supported by the Grant agency of the Czech republic under the grant P201/12/G028.
BC and FG are supported in part by the FWO-Vlaanderen through the project G.0114.10N.
FG is an Aspirant FWO-Vlaanderen.



\begin{thebibliography}{99}
  
\bibitem{Calabrese:2005in}
  P.~Calabrese, J.~L.~Cardy,
 \emph{``Evolution of entanglement entropy in one-dimensional systems,''}
\emph{J.\ Stat.\ Mech.\ }  {\bf 0504}, P04010 (2005) [ {\tt cond-mat/0503393}];
  
   P.~Calabrese, J.~Cardy,
  \emph{``Entanglement entropy and conformal field theory,''}
\emph{J.\ Phys.\ A} {\bf A42}, 504005 (2009)  [{\tt arXiv:0905.4013 [cond-mat.stat-mech]}].
  

\bibitem{Ryu:2006bv}
  S.~Ryu, T.~Takayanagi,
  \emph{``Holographic derivation of entanglement entropy from AdS/CFT,''}
 \emph{Phys.\ Rev.\ Lett.\ } {\bf 96}, 181602 (2006)
  [{\tt hep-th/0603001} ];
  
    S.~Ryu, T.~Takayanagi,
  \emph{``Aspects of Holographic Entanglement Entropy,''}
  \emph{JHEP} {\bf 0608}, 045 (2006)
  [{\tt hep-th/0605073}].
  
\bibitem{Hubeny:2007xt}
  V.~E.~Hubeny, M.~Rangamani, T.~Takayanagi,
  \emph{``A Covariant holographic entanglement entropy proposal,''}
  \emph{JHEP} {\bf 0707}, 062 (2007)
  [{\tt arXiv:0705.0016 [hep-th]}].
  
    
\bibitem{AbajoArrastia:2010yt}
  J.~Abajo-Arrastia, J.~Aparicio, E.~Lopez,
 \emph{``Holographic Evolution of Entanglement Entropy,''}
  \emph{JHEP} {\bf 1011 } (2010)  149.
  [{\tt arXiv:1006.4090 [hep-th]}].
  
\bibitem{Balasubramanian:2010ce}
  V.~Balasubramanian, A.~Bernamonti, J.~de Boer, N.~Copland, B.~Craps, E.~Keski-Vakkuri, B.~Muller, A.~Schafer {\it et al.},
  \emph{``Thermalization of Strongly Coupled Field Theories,''}
  \emph{Phys.\ Rev.\ Lett.\ } {\bf 106 } (2011)  191601.
  [{\tt arXiv:1012.4753 [hep-th]}]; 
  
  V.~Balasubramanian, A.~Bernamonti, J.~de Boer, N.~Copland, B.~Craps, E.~Keski-Vakkuri, B.~Muller and A.~Schafer {\it et al.},
  \emph{``Holographic Thermalization,''}
  \emph{Phys.\ Rev.\ D} {\bf 84} (2011) 026010
  [{\tt arXiv:1103.2683 [hep-th]}].
  
\bibitem{Headrick:2010zt}
  M.~Headrick,
  \emph{``Entanglement Renyi entropies in holographic theories,''}
\emph{Phys.\ Rev.\  }{\bf D82 } (2010)  126010.
  [{\tt arXiv:1006.0047 [hep-th]}].
  
\bibitem{Balasubramanian:2011at}
  V.~Balasubramanian, A.~Bernamonti, N.~Copland, B.~Craps and F.~Galli,
  \emph{``Thermalization of mutual and tripartite information in strongly coupled two dimensional conformal field theories,''}
  \emph{Phys.\ Rev.\ D} {\bf 84} (2011) 105017
  [{\tt arXiv:1110.0488 [hep-th]}].
  
\bibitem{Balasubramanian:2011wt}
  V.~Balasubramanian, M.~B.~McDermott and M.~Van Raamsdonk,
  \emph{``Momentum-space entanglement and renormalization in quantum field theory,''}
 \emph{Phys.\ Rev.\ D} {\bf 86} (2012) 045014
  [{\tt arXiv:1108.3568 [hep-th]}].
  
\bibitem{Hayden:2011ag}
  P.~Hayden, M.~Headrick, A.~Maloney,
  \emph{``Holographic Mutual Information is Monogamous,''}
    {\tt arXiv:1107.2940 [hep-th]}.

\bibitem{Tonni:2010pv}
  E.~Tonni,
  \emph{``Holographic entanglement entropy: near horizon geometry and disconnected regions,''}
\emph{JHEP} {\bf 1105 } (2011)  004.
  [{\tt arXiv:1011.0166 [hep-th]}].
  
  
\bibitem{alice}
A.~Bernamonti, 
\emph{``Applications of the AdS/CFT correspondence to strongly coupled dynamics,"}  PhD thesis,
 \emph{VUBPress} (2012).
 
 
\bibitem{Allais:2011ys}
  A.~Allais and E.~Tonni,
  \emph{``Holographic evolution of the mutual information,''}
\emph{JHEP} {\bf 1201} (2012) 102
  [{\tt arXiv:1110.1607 [hep-th]}].
  

\bibitem{Callan:2012ip}
  R.~Callan, J.~-Y.~He and M.~Headrick,
  \emph{``Strong subadditivity and the covariant holographic entanglement entropy formula,''}
 \emph{JHEP} {\bf 1206} (2012) 081
  [{\tt arXiv:1204.2309 [hep-th]}].

  
\bibitem{Wall:2012uf}
  A.~C.~Wall,
  \emph{``Maximin Surfaces, and the Strong Subadditivity of the Covariant Holographic Entanglement Entropy,''}
 {\tt arXiv:1211.3494 [hep-th].}
\end{thebibliography}
\end{document}